# Prediction of linear cationic antimicrobial peptides based on characteristics responsible for their interaction with the membranes

*Boris Vishnepolsky and Malak Pirtskhalava*

I. Beritashvili Center of Experimental Biomedicine, Tbilisi 0160, Georgia

**ABSTRACT**

The possibility to use antimicrobial peptides (AMP) for clinical purposes as a substitute for conventional antibiotics has become a cause of arising huge amount of studies which are focused on the features of AMP, their mechanism of action and the design of novel peptides. Of particular interest are *in silico* methods of research of AMP, that allow both to predict the antimicrobial activity of the peptides based on their sequence and to serve as the first step to design new antimicrobial peptides. Most available prediction methods use common approach for different classes of AMP. Contrary to available approaches, in this study we suggest, that a strategy of prediction should be based on the fact, that there are several kinds of AMPs which are vary in mechanisms of action, structure, mode of interaction with membrane etc. According to our suggestion for each kind of AMP a particular approach has to be developed in order to get high efficacy. Consequently in this paper a particular but the biggest class of AMP - linear cationic antimicrobial peptides (LCAP) - has been considered and a newly developed simple method of  LCAP prediction described.

The aim of this study is  the development of  a simple method of discrimination (based on threshold value only) of AMP from non-AMP, the efficiency of which will be determined by efficiencies of selected descriptors only and comparison the results of the discrimination procedure with the results obtained by more refined and complicated discriminative methods. As descriptors the physicochemical characteristics responsible for capability of the peptide to interact with an anionic membrane were considered. The following characteristics such as hydrophobicity, amphiphaticity, location of the peptide in relation to membrane, charge, propensity to disordered structure were studied. On the basis of these characteristics a new simple algorithm of prediction is developed and evaluation of efficacies of the characteristics as descriptors performed. The results of our study show that three descriptors: hydrophobic moment, charge and location of the peptide along the membranes can be used as discriminators of LCAPs. For the training set our method gives the same level of accuracy as more complicated machine learning approaches (SVM, RF, DA) offered as CAMP database service tools. For the test set sensitivity obtained by our method gives even higher value than the one obtained by CAMP prediction tools.

**INTRODUCTION**

Antimicrobial peptides (AMP) are small peptides of low length, which interact with the bacterial cells and kill them. The great interest in these proteins is explained by their possible use for clinical purposes as a substitute for conventional antibiotics when resistance takes place[1].

Most of AMP act directly on the bacterial membrane, consequently it is difficult for bacteria to develop immunity against antimicrobial peptides[2]. Recently, there have been a large number of both theoretical and experimental studies which were focused on the properties of AMP, their mechanism of action and the design of novel peptides (see for example review[3]). Of particular interest are *in silico* methods of research of AMP that allow both to predict the antimicrobial activity of the peptides based on their sequence and to serve as the first step to design new antimicrobial peptides. Methods for predicting AMP are based on some general properties which distinguish AMP from similar peptides which do not have antimicrobial activity.

Available prediction methods are generally based on discriminative analysis and (essentially) machine learning methods[4-12]. These methods as a positive training set have used a full set of antimicrobial peptide sequences, not taking into account variation in mechanisms of action, structure, mode of interaction with membrane and other differences. Contrary to available approaches, we think, that strategy of prediction should be based on the fact, that there are at least four kinds of AMPs for which four independent algorithms of prediction have to be developed in order to get high efficacy. For these four types of AMPs we can consider: cationic linier peptides (LCAP), cationic peptides stabilizing structure by inter-chain covalent bond (CCP), peptides rich in proline and arginine (PRP) and anionic antimicrobial peptides (AAP).

Cationic antimicrobial peptides (CAP) of LCAP type in addition to positive charge and amphiphilicity, possess simple mechanism of structure stabilization in membrane, hydrogen bonding only[13,14]. Absence of any other stabilization factors gives possibilities to determine the forces governing peptide-lipid or peptide-peptide interactions and predict structure of the peptides in water and membrane environment on a base of only sequence information. Consequently, quantitative characteristics for prediction would be easily revealed on the basis of sequence only. Structures of CAP of CCP type due to inter-chain bonds are more stable and structurally complicated both in water and in membrane environment. But despite the fact, that the forces governing peptide-membrane or peptide-peptide interactions are identical to the case of LCAP, complicated 3D structure and lack of information about 3D structure require a principally different approach for the development of CCP prediction algorithm. It's known that peptides of PRP type are penetrating. In other words, they don't destabilize membrane and as a rule have a target inside cell[15,16]. It's clear that the development of the algorithm for the prediction of CAP of PRP type requires peculiar approach. For AMP of AAP type mode of action principally differs from CAP and the development of the algorithm for prediction AAP indeed requires its own approach. In this work are considered CAP of LCAP type only. According to the available databases[17] this is the biggest class of antimicrobial peptides.

Prediction accuracy is largely determined by the set of descriptors which can be used in prediction. Most current methods use a large number of characteristics for AMP prediction, using their optimization by machine learning methods, such as ANN and SVM[4-12]. Meanwhile, the influence of the individual characteristics on the AMP prediction is studied much more weakly. In this paper, we describe the influence of the characteristics that may be responsible for the prediction of LCAP on the basis of their basic function - interaction with the bacterial membrane.

There are a large number of proteins that interact with the membrane also and so resemble AMP in this regard. For instance the so-called transmembrane proteins are generally inserted into the membrane but without destroying it. It is clear, that a selection pressure on sequence random variation directs evolution of peptides with particular function (for instance transmembrane protein fragments (TMP) , LCAP etc ). So, in order to determine what characteristics efficiently distinguish LCAP from other peptides (other membrane-interactive or non-functional (random)) we think that it's reasonable to make comparative analysis of sequences of the three sets of peptides: LCAP, TMP and randomly selected fragments from the soluble proteins (RFP). This work concerns just comparative analysis of LCAP, TMP and RFP sequences.

Consequently, an attempt to reveal that characteristics which can discriminate antimicrobial peptides from both soluble non-membrane proteins and transmembrane proteins (or fragments of membrane proteins) have been done. Taking into account the structure of the bacterial membrane, which is anionic lipid bilayer, amphiphatic in nature, it can be assumed that for discriminators the following characteristics are convenient: 1) hydrophobicity; 2) amphiphaticity 3) charge 4) propensity to the aggregation 5) propensity to disordering. We think that just the values of the these characteristics are responsible for: a) capability of the peptide to interact with an anionic membrane; b) the results of interaction (mechanisms of action).

Quantitative estimation of all the characteristics except amphiphaticity requires information on amino acid sequences of the peptides only. Amphiphaticity in addition needs three dimensional structure information. The exact three dimensional structure of most antimicrobial peptides is unknown. But energetically the most favorable conformations for LCAP in membrane environment should presumably be saturated with hydrogen bonds[13,14]. So we are motivated to make estimation in regular structure approximation.

There are various statistical approaches for the prediction of AMP which take into account a number of different characteristics[4,5,11,18-23]. In this paper our goal is: a) to develop the simplest method of discrimination (based on threshold value only) of AMP from non-AMP, efficiency of which will be determined by efficiencies of selected descriptors only and b) to compare the results of the discrimination procedure with the results obtained by more refined and complicated discriminative methods such as SVM, ANN etc.

**RESULTS AND DISCUSSION**

**Optimization of the Descriptors**

The following descriptors were considered: hydrophobic moment, charge, location of the peptide in relation to membrane (LPM), linear hydrophobic moment and disordering.

The optimization of the most descriptors (except LPM) has been made by MOC criterion (see Experimental Section). The corresponding data are given in Table 1. Taking into account the fact that in the case of normal (random) distribution probability that z-score>2 equals 0.02, we can say that for all descriptors the observed probabilities that z-score>2 are higher than expected for normal distribution. It means we can assume some kind of selection pressure on sequence random variation.

**Table 1.** Optimal parameters for different descriptors

|  | Hydrophobicity scale | Fragment Length | Angle $\vartheta$ ° | MOC[5] No | % | d | θ | AUC_R | AUC_T |
|---|---|---|---|---|---|---|---|---|---|
| HM[1] | MF | 24 | 96 | 679 | 62.70 |  |  |  |  |
| CHARGE |  |  |  | 145 | 13.34 |  |  |  |  |
| HYDR[2] | KD | 21 |  | 258 | 22.30 |  |  |  |  |
| LM[3] | EG | 31 |  | 119 | 10.00 |  |  |  |  |
| DISORD[4] |  |  |  | 46 | 4.25 |  |  |  |  |
| LPM |  | 17 |  |  |  | 12.9 | 81.0 | 0.76 | 0.78 |

[1]Hidrophobic moment, [2]Hydrophobicity, [3]Linear hydrophobic moment, [4]Disordering
[5]A Number (No) and percent (%) of the peptides from LCAP set which satisfy MOC criterion.

For hydrophobic moment, for example, the optimal value of the turn ($\vartheta$) of residue ($\vartheta$ varied from 60° to180°) in regular structure is 96°, which shows that optimal secondary structure in LCAP set is $\alpha$-helix. This result can be expected.

For LPM was used another criterion of optimality (different from MOC) which was based on d and $\theta$ distributions in the considerable set (see Experimental Section). Assuming the peptide is $\alpha$-helical (see above) for each peptide we can calculate the energetically most favorable location (d and $\theta$) of the peptide fragments of the particular length. Fig 1 shows plot of relative density of the orientation and the depth of energetically most favorable fragments of length 17 on the basis of the three considered sets (LCAP, RFP10 and TMP).

From Figure 1 it is clear that the orientation distribution of peptides in different sets varies from each other. For the most LCAP the more energetically favorable depth is within 8-15 Å, which corresponds to the boundary of the membrane interface and the hydrophobic core (Fig. 1a). It also shows that most of the peptides are located at a relatively small angle to the membrane surface ($\theta$ ~90°). These results are consistent with experimental data, according to which most of the CAP penetrate into the membrane at a shallow depth parallel to the membrane surface[24]. Maximum density on the (d, $\theta$) plot for LCAP set is higher than for the other peptide sets (RFP10 and TMP). From Fig. 1b it can be seen that it is energetically more favorable for the membrane proteins to penetrate more deeply into the membrane (d=2-10 Å). At the same time peptides from the dataset of random protein fragments are located closer to the membrane surface (d=12-30 Å; see Fig. 1c). Peptides from the later set are distributed on the (d, $\theta$) plot less densely than from the other peptide sets.

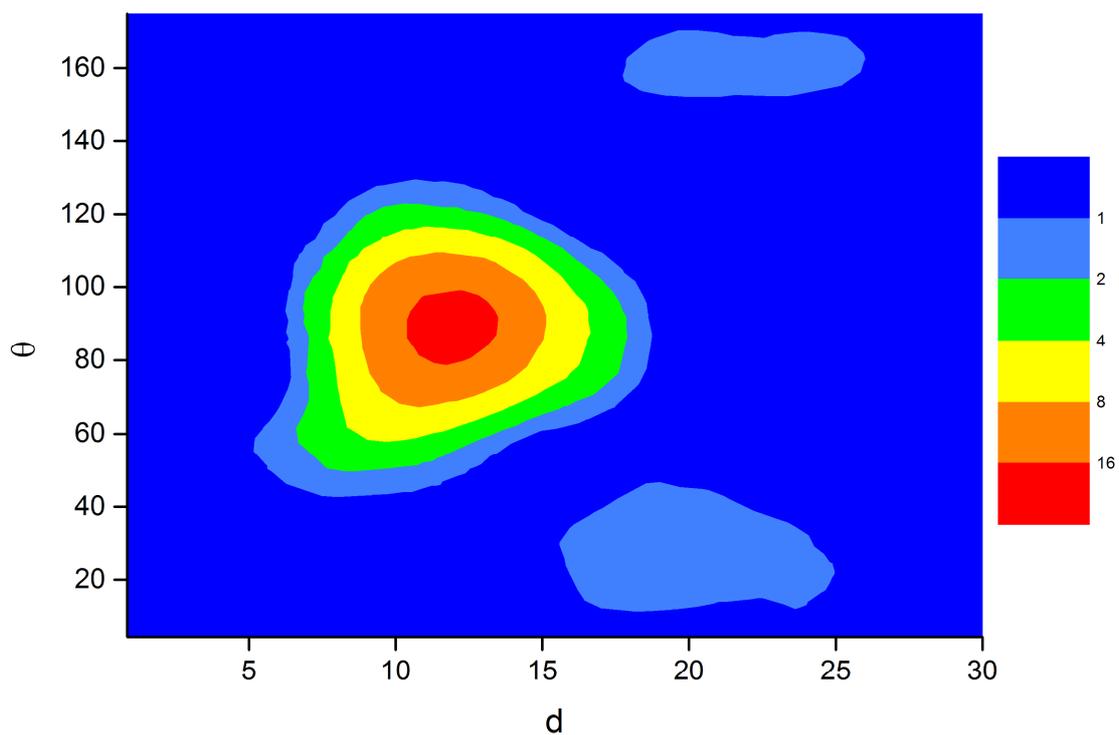

**Fig. 1a.** Plot of relative density of the orientation (θ) and the depth (d) of energetically most favorable fragments of length 17 on the basis of the LCAP set. The values of the density are given relative to the densities of uniform distribution.

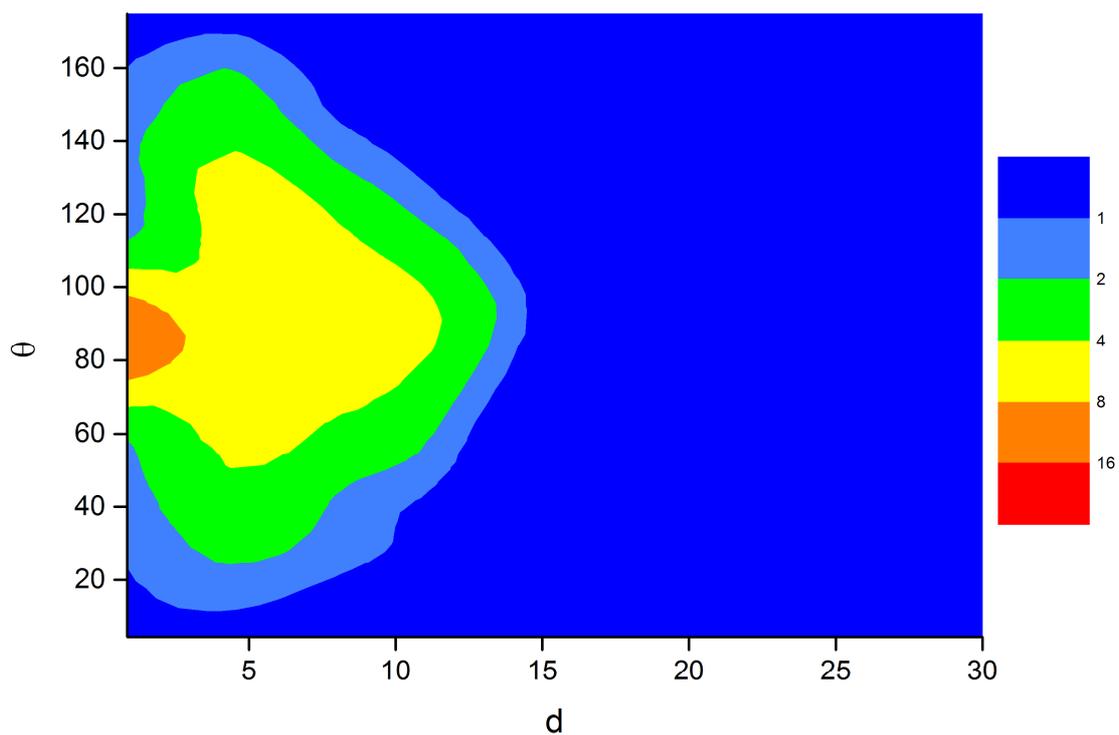

**Fig. 1b.** Plot of relative density of the orientation (θ) and the depth (d) of energetically most favorable fragments of length 17 on the basis of the TMP set. The values of the density are given relative to the densities of uniform distribution.

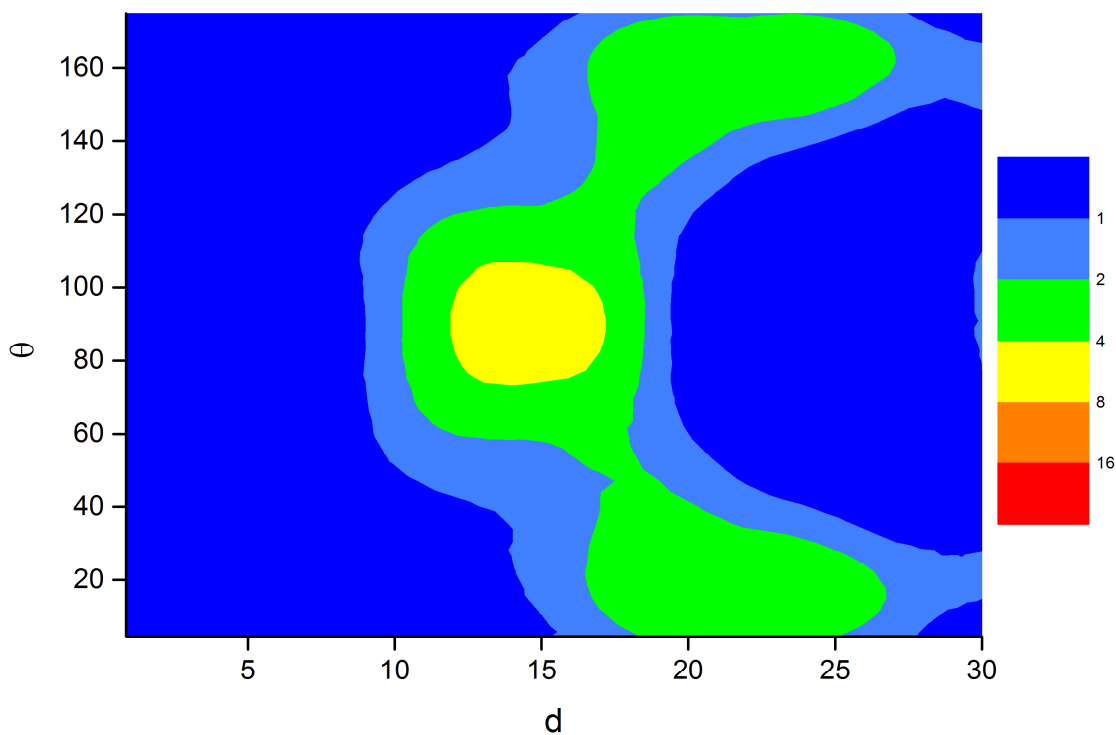

**Fig. 1c.** Plot of relative density of the orientation (θ) and the depth (d) of energetically most favorable fragments of length 17 on the basis of the RFP500 set. The values of the density are given relative to the densities of uniform distribution.

Based on these data we have decided to use receiver operating characteristic (ROC) curve, to quantify differences of LCAP from membrane proteins and soluble proteins fragments (see Experimental Section). Calculations have shown that the optimal values of the penetration depth and angle (d and θ) are 12.9 Å and 81° at fragment length of 17 amino acids. The optimal AUC_R and AUC_T values (see Experimental Section) for the two datasets (RFP500 and TMP) at the same time are 0.76 and 0.78 respectively. So we concluded that LPM can be used as a descriptor to distinguish LCAP from other peptides.

**Evaluation of the efficiency of LCAP prediction**

Receiver operating characteristic (ROC) curves were used to evaluate the effectiveness of various characteristics. ROC curves, plotted for each characteristic are shown in Fig. 2-8. Quantitative evaluation of effectiveness of the characteristics is made on the basis of the following quantities: area under the ROC curve (defined as AUC_R relative to the RFP500 negative set and defined as AUC_T relative to the TMP negative set); a threshold for each characteristic by which prediction of peptide antimicrobiality will be done. A threshold is determined by a point on the ROC curve closest to the point (0,1). Sensitivity, specificity and balanced accuracy for the thresholds have been evaluated.

**Table 2.** Comparison of the different descriptors for training set (LCAP and RFP500)

|  | AUC_R*100 | AUC_T*100 | $R_{min}$[5]*100 | $S_n$*100 | $S_p$*100 | BAC*100 |
|---|---|---|---|---|---|---|
| HM[1] | 88.63 | 92.01 | 23.32 | 80.79 | 86.77 | 83.78 |
| CHARGE | 81.57 | 92.93 | 35.02 | 83.29 | 69.21 | 76.25 |
| LPM | 76.12 | 78.32 | 37.38 | 76.36 | 68.12 | 72.24 |
| HYDR[2] | 71.18 | 11.09 | 47.96 | 64.27 | 69.15 | 66.14 |
| LM[3] | 56.09 | 33.37 | 68.65 | 40.63 | 65.54 | 53.08 |
| DISORD[4] | 46.59 | 93.75 | 74.96 | 50.97 | 43.30 | 47.13 |

[1]Hidrophobic moment, [2]Hydrophobicity, [3]Linear hydrophobic moment, [4]Disordering

[5]Distance from the point (0,1) to the point on the ROC curve closest to the point (0,1)

Condition of z-score>$z_i$ or z-score<$z_i$ was used to calculate i-th point of the ROC curve, that is the values of sensitivity and specificity ($S_{ni}$ and $S_{pi}$) (see Experimental Section). Condition of z-score>$z_i$ or z-score<$z_i$ was proposed based on the assumption that the values of the considered descriptors must be higher (for hydrophobic moment, charge and linear hydrophobic moment) or less (for hydrophobicity and disordering) for LCAP than for non-AMP. If the assumption is true AUC for each descriptor will be higher than 0.5. It can be noted that the higher the value of AUC the better the descriptor discriminates AMP from non-AMP. The value of AUC for good descriptors must be no less than 0.7. But as we can see our results show that the values of AUC_R for linear moment is close to 0.5 (see Table 2 and Fig. 2 ) and for disordered even less than 0.5 (see Table 2 and Fig. 3). It means that the last characteristics cannot distinguish antimicrobial from non-antimicrobial peptides. The low value of AUC_R=0.56 for the linear moment suggests that for the most antimicrobial peptides there is no significant linear separation of hydrophobic and hydrophilic residues along the peptide chain. On the other hand, ROC curve plotted for linear moment of TMP set relative to the negative set RFP500 (ROC_RM (Fig. 4)) gives the value 0.73 for the area under the ROC curve (AUC_RM =0.73). These differences between the values of AUC_R and AUC_RM can be explained by the fact that in contrast to antimicrobial, in the transmembrane peptides linear separation of hydrophobic and hydrophilic group of residues occurs. Such separation was revealed by other authors[25,26] also, which supposed that amphyphilic residues are concentrated at the ends of the transmembrane helix while hydrophobic residues are located in the middle.

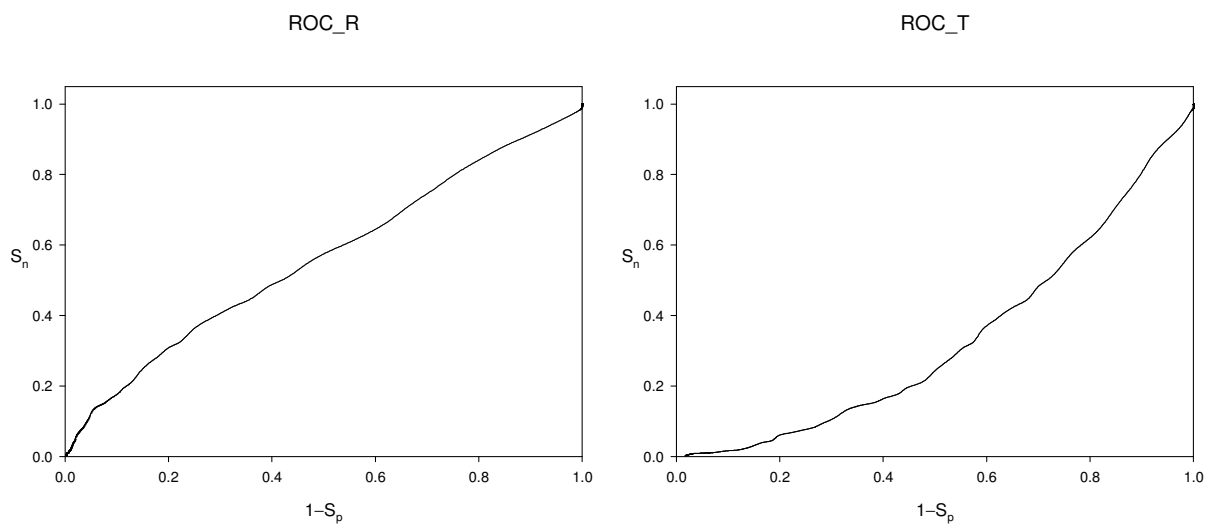

**Fig. 2.** ROC curves for evaluation prediction quality of linear hydrophobic moment for training set (ROC_R) for LCAP and RFP500 sets and (ROC_T) for LCAP and TMP sets

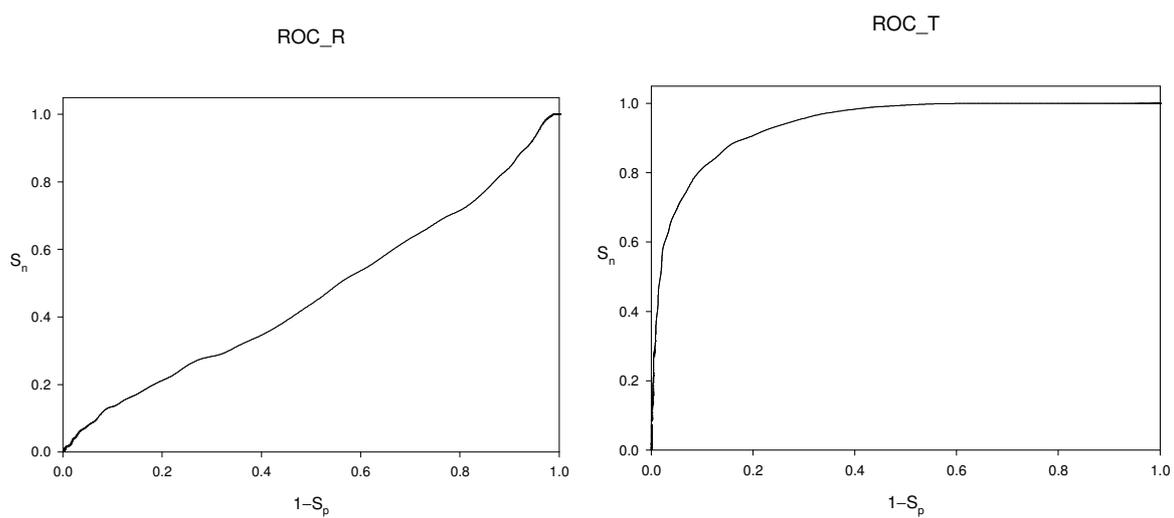

**Fig. 3.** ROC curves for evaluation prediction quality of disordering for training set (ROC_R )for LCAP and RFP500 sets and (ROC_T) for LCAP and TMP sets

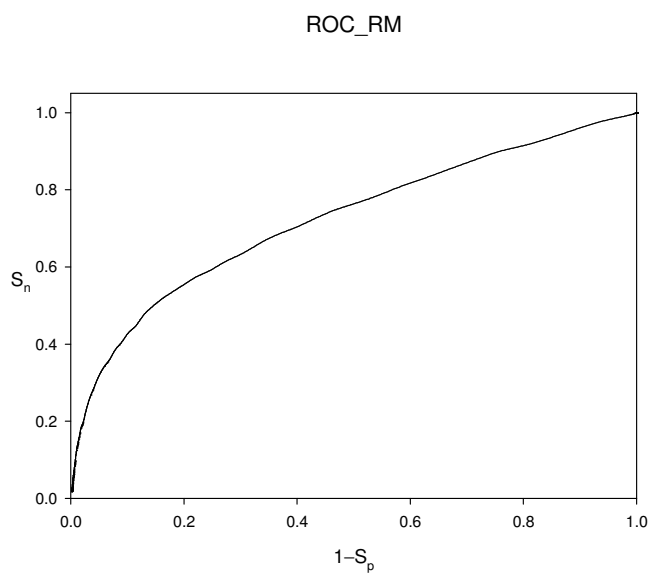

**Fig. 4.** ROC curve for evaluation prediction quality of the transemembrane helixes of linear hydrophobic moment (ROC_R M) for TMP and RFP500 sets

AUC_R value for disordered is 0.47, which is less than 0.5, and it can be said that according to the proposed by Uversky criteria[27] antimicrobial peptides are more ordering than fragments of random proteins. This may be due to the fact that the Uversky criterion, which determines the degree of the disorder of globular proteins is not suitable for the evaluation of the disorder of small peptides.

In the case of hydrophobicity, hydrophobic moment, LPM and charge AUC_R> 0.7 (see Table 2 and Fig. 5-8), which indicate that these characteristics can be used to distinguish antimicrobial from soluble non-antimicrobial peptides. For the hydrophobic moment and charge AUC_T > AUC_R. Based on this we can suggest that if the value of the last characteristics can discriminate a peptide (potential LCAP) from the non-membrane peptides it should rather discriminate the peptide from the transmembrane peptides. So for these characteristics only single threshold defined from ROC_R can be used, because for this threshold sensitivities are the same for ROC_R and ROC_T, but specificity and so accuracy obtained from ROC_T is larger than from ROC_R and LCAP peptides can be discriminated from the membrane peptides with accuracy obtained from the threshold defined from ROC_R only.

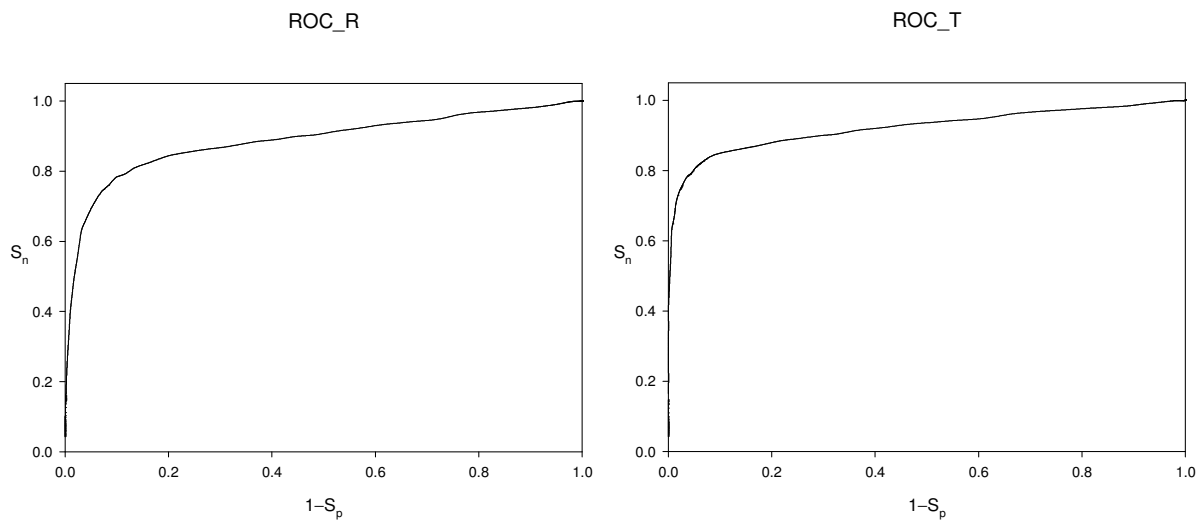

**Fig. 5.** ROC curves for evaluation prediction quality of hydrophobic moment for training set (ROC_R) for LCAP and RFP500 sets and (ROC_T) for LCAP and TMP sets

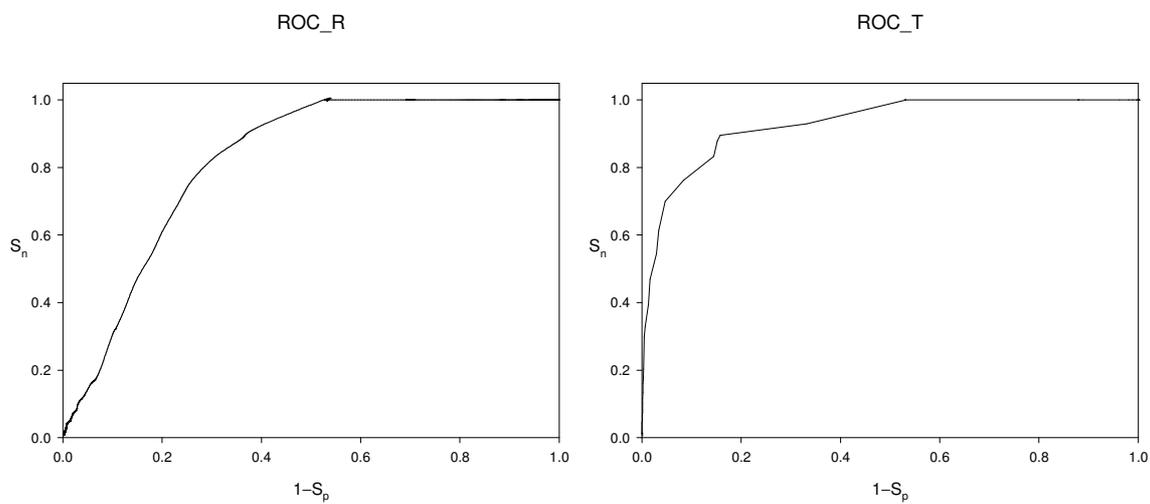

**Fig. 6.** ROC curves for the evaluation of prediction quality of charge for training set (ROC_R) for LCAP and RFP500 sets and (ROC_T) for LCAP and TMP sets

LPM was already optimized in such a way that the greater difference from the RFP500 and TMP sets was reached (see above). AUC_R for this descriptor is 0.76, so it can be used as LCAP characteristic (see Table 2).

Though for the hydrophobicity AUC_R=0.71, but AUC_T =0.11<0.5 (see Table 2), it means that the LCAP has a lower average hydrophobicity than the transmembrane helices, but greater than random fragments from the soluble proteins. Therefore, we cannot use single threshold to discriminate non-antimicrobial and antimicrobial peptides. We have not used this descriptor as discriminated LCAP characteristic.

The highest value of AUC_R corresponds to hydrophobic moment (AUC = 0.89) (Table 2). Thus we can suggest that this characteristic is the best separator between non-antimicrobial and antimicrobial peptides.

So three descriptors: hydrophobic moment, charge and LPM were selected as most effective LCAP descriptors. Given the above it can be assumed that the combined use of these three characteristics can improve the prediction of LCAP. To combine these characteristics we have taken into account the fact that for the separation of LCAP and non-AMP, specific threshold

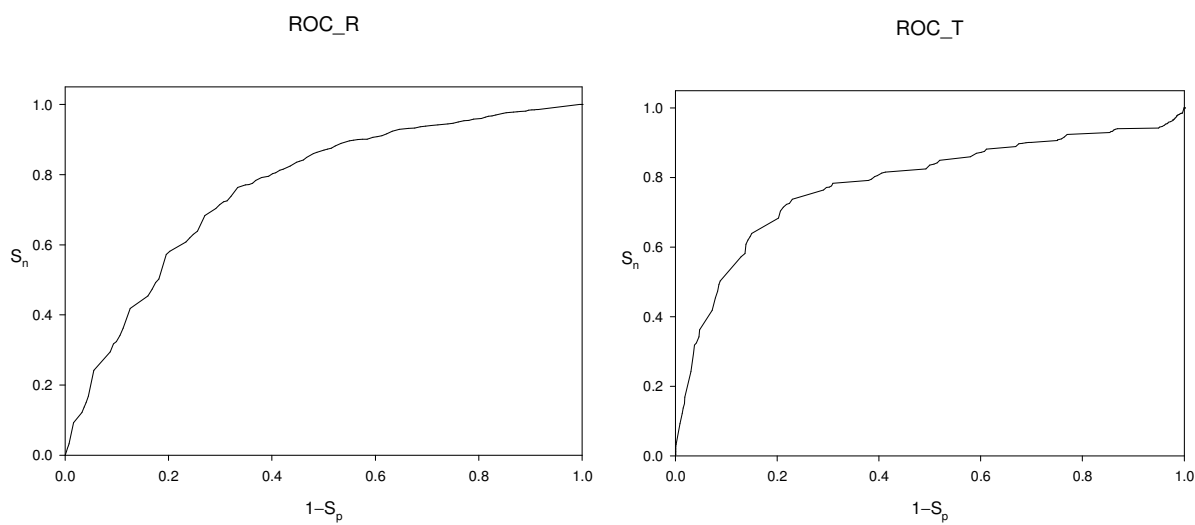

**Fig. 7.** ROC curves for evaluation prediction quality of location of the peptide along the membrane (LPM) for training set (ROC_R) for LCAP and RFP500 sets and (ROC_T) for LCAP and TMP sets

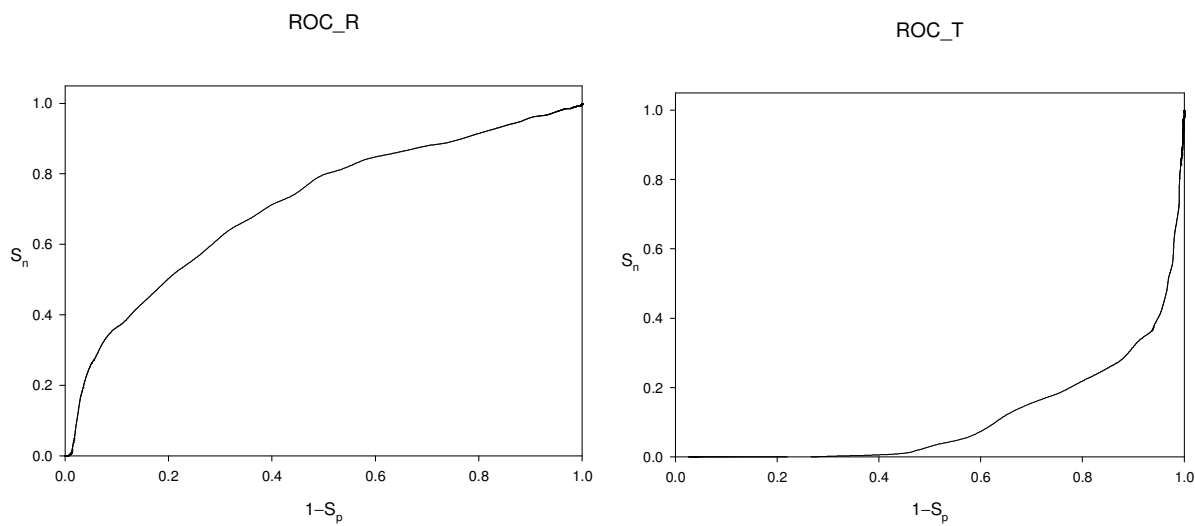

**Fig. 8.** ROC curves for evaluation prediction quality of hydrophobicity for training set (ROC_R) for LCAP and RFP500 sets and (ROC_T) for LCAP and TMP sets

that can be obtained from the analysis of ROC curves were used. Accordingly, by changing synchronously these thresholds for different characteristics we can simply optimize these thresholds to obtain the greatest accuracy. The corresponding balanced accuracy for the hydrophobic moment alone, hydrophobic moment and charge together and for the three characteristics together are 83.8, 84.6, 86.7, respectively (Table 2 and 3).

**Table 3.** Prediction quality of combined use of three descriptors for training set (LCAP and RFP500)

|  | AUC_R*100 | $S_n$*100 | $S_p$*100 | BAC*100 |
|---|---|---|---|---|
| HM[1]+CHARGE | 87.35 | 86.89 | 82.31 | 84.60 |
| HM[1]+CHARGE+LPM | 89.25 | 81.44 | 92.02 | 86.73 |

[1]Hidrophobic moment

We have also tried to evaluate our results with other prediction methods. As we have mentioned above, several computational methods[4,5,11,18-23] have been proposed for the predicting AMPs. However, methods[4,5,18] did not contain available web services for testing our datasets. BACTIBASE[19,20] and PhytAMP[21] methods were specifically designed for bacteriocin and plant respectively. As for AntiBP[22] and AntiBP2 methods[23], they were designed to identify the AMPs in a protein sequence, and hence could not be used to compare with our method. So to make the comparison meaningful, our method was compared with CAMP method[11], which was developed based on the Random Forests (RF), SVM, and Discriminant Analysis (DA). This method can be used for the evaluation of the sensitivity, specificity and accuracy for the considered positive set. As a set of non-antimicrobial peptides (negative set) we have used a set of 10 peptide fragments (instead of 500) for each peptide in the AMP set (RFP10). The corresponding balanced accuracy for three considered characteristics together when using this set is 87.7. For the same positive and negative sets CAMP method gives the following values for the balanced accuracy : Random Forests (RF) - 90.7, SVM - 88.5, and Discriminant Analysis (DA) - 86.5 (Table 4).

**Table 4**. Prediction quality for different methods for sets LCAP and RFP10

|  | $S_n*100$ | $S_p*100$ | BAC*100 | AC*100 |
|---|---|---|---|---|
| HM[1] | 80.79 | 89.54 | 85.17 | 88.74 |
| HM[1]+CHARGE | 84.21 | 90.11 | 87.16 | 89.57 |
| HM[1]+CHARGE+LPM | 81.44 | 93.72 | 87.58 | 92.60 |
| SVM | 86.80 | 90.27 | 88.54 | 89.96 |
| RF | 89.57 | 91.80 | 90.69 | 91.59 |
| DA | 85.60 | 87.37 | 86.49 | 87.21 |

[1]Hidrophobic moment

For testing TPS positive set, i.e. set of 98 sequences selected from CAMP[11] predicted dataset (see Experimental Section) was used. The results of comparison for test set are shown in the Table 5.

We cannot use any additional test sets for non-AMP (negative set), so only sensitivity data were evaluated. The prediction of sensitivity for our method was 80.61, higher than for a CAMP. Our results show that prediction quality in case of using two or three descriptors together did not give much better results than when only hydrophobic moment was employed. For the test set the combined use of these descriptors even impair prediction quality. The main reason for it is that we have used very simple method for combining these descriptors. Our results on the training set show that the prediction quality of the LCAP based on one descriptor, hydrophobic moment, does not very strongly differ from the CAMP method, moreover in case of the test set outperforms CAMP. We want to emphasize the fact, that CAMP method uses a combination of numerous characteristics and more complicated, refined and effective discriminative methods[11]. High performance of our approach can be explained by the fact that we have used only one class of AMP, cationic linier peptides. Our results confirm the assumption that prediction of AMP is preferable to make for the peculiar class separately, using particular approach in each case.

**Table 5.** Prediction quality for different methods for test set

|  | $S_n*100$ |
|---|---|
| HM[1] | 80.61 |
| HM[1]+CHARGE | 80.61 |
| HM[1]+CHARGE+LPM | 79.59 |
| SVM | 74.49 |
| RF | 76.53 |
| DA | 75.51 |

[1]Hidrophobic moment

Another descriptor that we do not study here is propensity to the aggregation. The question of AMP aggregation is difficult and unclear. There is speculation that AMP greatly differ in the predisposition to aggregation[28]. A wide range of proposed mechanisms of AMP action can be explained by the fact that AMP behave differently in terms of the stability of their aggregates both in the membrane and in the aqueous environment. However, there are papers that suggest that propensity to the aggregation based characteristics are good descriptors[4]. But our assessment, not presented here for a number of reasons, indicates the opposite. Therefore, in this article we do not consider this descriptor, preferring to make more thorough (full) estimates and submit them in the next paper.

**EXPERIMENTAL SECTION**

**Benchmarks**

**Training sets.** For the analysis of the characteristics the following benchmarks were selected: set for LCAP, set for randomly selected fragments from the soluble proteins and set of fragments from transmembrane proteins. The LCAP set was selected from APD2 database[17] and consists of 1083 peptides (positive set). To estimate the discriminative efficiency of characteristics a set of non-antimicrobial peptides has been required. Because there are small number of peptides with experimentally-verified no antimicrobial activity[5], we have used a voluminous set of random sequences, in other words a set of sequences with a great variety of functions. So the last set can be considered as non-functional set on average, as well as non-antimicrobial (negative set). The set of random sequences was selected from an UniProt using the filters : non- AMP, non-membrane and non secretory proteins. There were used three such sets. The first set (RFP10000) was used for optimizing parameters for various descriptors and consists of randomly chosen fragment in amount of 10000 for each length of peptides from 4 to 50 amino acids. The other sets were used for the estimation of the descriptors by ROC curves. 500 (for RFP500) and 10 (for RFP10) randomly selected fragments from globular proteins with length corresponding to each peptide from LCAP set have been included into these sets. The last set was used for comparing our results with other available prediction tools.

For membrane proteins a full set of transmembrane (helices) fragments of more than 11 residues from database of transmembrane proteins PDB-TM[29,30] was choosen. This set contains 1691 sequences (TMP set).

**Test set.** Test set is compiled on the basis of CAMP[11] predicted dataset contained 1,153 sequences identified as antimicrobial based on the evidences of similarity or annotations in NCBI as 'antimicrobial regions', without experimental evidences. After eliminating sequences: containing non-standard residues 'B', 'J', 'O', 'U', 'X', or 'Z'; disulfide bonds; having full negative charges; with the length of more than 50 amino acids and rich in Pro and Arg, only 98 sequences were left (TPS). TPS will serve as independent positive test dataset. As mentioned above, we could not use any additional independent sample as independent negative test set.

**Optimization of the parameters defining the characteristics of AMP**

There is evidence, especially for disulphide-bounded AMP, that despite their short length they are unions of functional (structural) blocs[31]. So we can propose, that linear peptides are arranged in bloc principle also and not all the considered peptide, but part of it can participate in the interaction with the membrane. Accordingly, for each peptide, the descriptors were calculated for all fragments (windows) of a certain length and peptides are characterized by the particular fragment selected on certain criteria.

The values of different descriptors, in most cases depend on various parameters, such as length of the fragment for which considered characteristic for the peptide is computed, hydrophobicity scale (see below) etc. It is necessary to choose optimal parameters for descriptors on the base of LCAP set. Optimization of the descriptors was made by the requirement of increasing the ratio (percent) of the peptides for which the probability of appearance of their sequence as a result of random normal process is less than P. The value of P was determined by z-score. Main criterion of optimality (MOC) was the maximality of the number of peptides from the LCAP set having z-score >2. Exception was a location of the peptide in relation to membrane, for which optimization has been made differently ( see below).

**Hydrophobicity**

The AMP overall hydrophobicity defined as the sum of transfer (from water into the hydrophobic environment) energy of the residue (hydrophobicity) can be used as AMP characteristic. In the literature, there are a large number of papers[32-37] which define transfer energies of the amino acids (hydrophobicity scales). The values of the transfer energies in these scales depend on the method of determination and differ from scale to scale. Therefore, the hydrophobicity scale can be used as an optimization parameter for assessing the suitability of the hydrophobicity as AMP characteristics. The following hydrophobicity scales were considered: KD[32], WW[33], UHC[34], Hes[35], EG[36] and MF[37].

For each peptide, hydrophobicity was calculated for all fragments of a certain length and the peptide characteristic was defined by the fragment of the highest hydrophobicity. Therefore peptide fragment length can be the other optimization parameter. The optimal length and hydrophobicity scale were chosen by MOC. Fragment length was varied within the range of 4-50 residues. Moreover, if the peptide length is less than the length of the considered fragment hydrophobicity was computed for the full peptide. A similar method for optimizing the fragment length was used for the other descriptors.

**Amphipathicity**

One of the main features of antimicrobial peptides is their amphipathicity[38]. The separation of hydrophobic and hydrophilic regions in these peptides can be realized in one of the two ways: due to the internal 3D structure and by the linear separation that is due to the uneven distribution of hydrophobic and hydrophilic residues along the peptide chain. Accordingly, two characteristics were used for the evaluation of amphipathicity: hydrophobic moment[39] and linear hydrophobic moment (see below).

**Hydrophobic Moment.** Hydrophobic moment was estimated by Eisenberg[39].

$$\mu = \left( \left[ \sum_{n=1}^{N} h_n \sin(\vartheta n) \right]^2 + \left[ \sum_{n=1}^{N} h_n \cos(\vartheta n) \right]^2 \right)^{1/2}$$

where $\mu$ is hydrophobic moment of the peptide, contained $N$ amino acids, $h_n$ is the numerical hydrophobicity of the $n$ th residue, and $\vartheta$ is turn of the residue along the helix axis.

According to the formula the existence of regular conformation is assumed. As mentioned above LCAP in membrane environment is likely to have regular secondary structure. So $\vartheta$ is used as a parameter which determines hydrophobic moment. The last parameter was used as optimization parameter and varied from 60 to 180°. Hydrophobicity scale and fragment length were also used as optimization parameters. Optimization of the parameters was carried out by MOC.

**Linear hydrophobic moment.** As mentioned above separation of the hydrophobic and hydrophilic parts may also be carried out due to an uneven distribution of hydrophobic and hydrophilic residues along the peptide chain. In order to estimate the separation along the chain, we have introduced the characteristic "linear hydrophobic moment", which is defined as follows:

$$M = D(\sum h_k^+ - \sum h_k^-),$$

where

$$D = |\sum h_k^+ \cdot k / \sum h_k^+ - \sum h_k^- \cdot k / \sum h_k^-|$$

D is the distance between the centers of hydrophobic and hydrophilic parts of the considered fragment of length N; $k=1,N$, $h_k^+$ and $h_k^-$ - transfer energy of the k-th residue from water to the hydrophobic environment under the conditions that $h_k^+ > 0$ corresponds to hydrophobic residue and $h_k^- < 0$ corresponds to hydrophilic residue.

Hydrophobicity scale and fragment length were used as optimization parameters. Optimization of the parameters was carried out by MOC.

**Charge**

Cationic antimicrobial peptides at neutral pH have a positive charge due to the large percentage of Lys and Arg, which facilitates them to interact with the negatively-charged membrane. So it is natural to assume that the charge of the peptide can be considered as a characteristic of LCAP. Because electrostatic interaction is long-term, we think that the net charge of the whole peptides determines the results of interaction with membrane. So the charge was calculated for the entire peptide.

**Location of the peptide in relation to membrane (LPM)**

Mechanism of action of AMP largely depends on their energetically most favorable location within the membrane bilayer. Taking into account the fact, that majority of the LCAP peptides have α-helical conformation in the membrane environment (see above and Results section) LPM was described by the penetration depth (d) i.e. distance of the geometrical centre of peptide helix from membrane surface and angle (θ) between peptide helix axis and perpendicular to the membrane surface. It would be interesting to explore the possibility of using d and θ as discriminators to distinguish antimicrobial from non-antimicrobial peptides. In contrast to the previous LCAP characteristics, location of the LCAP within the bilayer is an integrated feature that will largely depend on the other previously considered characteristics (hydrophobic, amphipathicity, charge). To calculate d and θ, the hydrophobic potential designed by Senes et all[40], which represents the energy difference between the residue in water and within the bilayer at a given depth, is used. All calculations of LPM were performed for different fragment(window) lengths and the peptide characteristic was defined by the fragment with minimal energy.

Another (different from MOC) approach was used for the optimization of d and θ. The approach is based on receiver operating characteristic (ROC) curve analysis. ROC curve, which represents the dependence of sensitivity ($S_n$) (y-axis) versus 1 - specificity ($S_p$) (x-axis) was used for quantifying differences of LCAP from membrane proteins and soluble protein fragments. RFP500 and TMP were used as the negative sets. For each LPM was calculated area under the ROC curve AUC, defined as AUC_R relative to the RFP500 negative set and as AUC_T relative to the TMP negative set. As above mentioned peptide fragment length varied and so used as optimization parameter. The maximization of (AUC_R + AUC_T) value was used to optimize LPM (d and θ). During optimization d and θ vary from 0-30 Å and 0-180° respectively and were used as optimization parameters for LPM. Other variables $\delta(d_k)$ and $\delta(\theta_k)$ (for each k-th d and θ) were used for plotting ROC curve also. For each of the values $d_k$ and $\theta_k$, $\delta(d_k)$ and $\delta(\theta_k)$ varied and for i-th their values $\delta(d_k)_i$ and $\delta(\theta_k)_i$, intervals $d_k \pm \delta(d_k)_i$, $\theta_k \pm \delta(\theta_k)_i$ i.e. i-th area on the (d, θ) plane is determined. The number of peptides from the positive datasets with energetically most favorable depth and orientation lying within the interval $d_k \pm \delta(d_k)_i$, $\theta_k \pm \delta(\theta_k)_i$ determines sensitivity $Sn_i$ and the number of peptides from the negative datasets with energetically most favorable depth and orientation lying within the same interval ($d_k \pm \delta(d_k)_i$, $\theta_k \pm \delta(\theta_k)_i$) determines specificity $Sp_i$. $Sn_i$ and $Sp_i$ give i-th point of the ROC curve. For each length, $d_k$ and $\theta_k$ ROC curves and consequently (AUC_R + AUC_T) values were calculated and maximum among the calculated values correspond to optimums of length, $d_k$ and $\theta_k$.

**Disordering**

Most of AMP have no specific structure in solution and only take a structure in the hydrophobic membrane environment. Therefore, it would be interesting to test the possibility to use disordering as a descriptor. To quantify the extent of disordering we used the formula proposed by Uversky at all.[27]. According to this formula the degree of disordering of globular protein under physiological conditions is defined by the relation

$$S = 2.785 <H> - 1.151 - <R>$$

$<H>$ the average hydrophobicity of the protein and $<R>$ its charge:
Negative values of S correspond to the protein to be disorded.

**Evaluation of the efficiency of characteristics**

Receiver operating characteristic (ROC) curves were used to evaluate the effectiveness of various characteristics. Each point i of the ROC curve corresponds to values of sensitivity and specificity ($Sn_i$ and $Sp_i$) which are calculated for the variable $z_i$. The number of peptides from the positive dataset (LCAP) with z-score>$z_i$ (for hydrophobic moment, charge and linear hydrophobic moment) and z-score<$z_i$ for (hydrophobicity and disordering) determines $Sn_i$ and the number of peptides from the negative datasets (RFP500 for ROC_R and TMP for ROC_T) with z-score>$z_i$ (for hydrophobic moment, charge and linear hydrophobic moment) and z-

score<$z_i$ (for hydrophobicity and disordering) determines $Sp_i$. Quantitative evaluations of the effectiveness of the characteristics are made on the basis of area under the ROC curve.

**Evaluation of the prediction quality**

A threshold for each characteristic was evaluated and the prediction of antimicrobiality of peptide was done on the basis of it. The threshold is determined by a point on the ROC curve closest to the point (0,1). Sensitivity, specificity and accuracy for the thresholds have been evaluated.

So the following equations were used to reflect the prediction quality:

$$S_n = TP/(TP+FN)$$

$$S_p = TN/(TN+FP)$$

$$BAC = (S_n + S_p)/2$$

$$AC = (TP+TN)/(TP+FN+TN+FP)$$

where $S_n$ reflects the sensitivity, $S_p$ the specificity, BAC the balanced accuracy, AC the accuracy.

The calculation of balanced accuracy is used for the evaluation of the prediction quality because the negative sets contain much more peptides than the positive ones and the balanced accuracy reflects equal influence of positive and negative sets irrespective of the number of contained peptides in them.

**ACKNOWLEDGEMENT**

The designated work has fulfilled by financial support of the CNRS / SRNSF Grant No 09/10.